\documentclass[prb,twocolumn,floatfix]{revtex4-1}

\usepackage{bm}
\usepackage{lipsum}
\usepackage{subfigure}
\usepackage[%
  backref=page,%
  pdfpagelabels=true,%
  plainpages=false,%
  colorlinks=false,%
  bookmarks=true,%
  hidelinks=true,
  pdfview=FitB]{hyperref}

\usepackage{latexsym}
\usepackage{graphicx}
\usepackage{longtable}
\usepackage{booktabs}

\newcommand{\eexp}{{\text{e}}}

\newcommand{\Pcal}{{\mathcal{P}}}

\newcommand{\vmu}{\bm{\mu}}

\newcommand{\ee}{{\text{e}}}  
\newcommand{\R}{{\textsuperscript{\textregistered}}}

\usepackage{float}
\restylefloat{table}

\usepackage{array}
\usepackage{verbatim}
\usepackage{amsmath}
\usepackage{color}
\usepackage{xcolor}
\usepackage{soul}
\usepackage{tabu}
\usepackage{multirow}
\usepackage[normalem]{ulem}
\usepackage[arrowdel]{physics}
\usepackage{listings}
\begin{document}
\graphicspath{{fig/}{./}}
\title{rt-TDDFT modeling of thermal emission by laser-heated glasses}
\author{Grigory Kolesov}
 \affiliation{
  Corning Inc., Painted Post, New York, USA
  }
\date{\today}

\begin{abstract}

In the laser processing of glass, a \~50-1000 $\mu$m-thick layer of glass  is
heated to a high temperature by the laser beam. Due to the shallow depth of this
hot layer, the infrared emission and absorption spectra may deviate from the black-body
spectra and can be influenced by the vibrational structure of the material.
Real-time time-dependent density functional theory (rt-TDDFT) modeling of the
thermal radiation by such hot layers allows us to calculate the emissivity and
thus to evaluate the reliability of the measurements conducted with thermal
cameras at specific wavelengths.

\end{abstract}

\maketitle

\section*{Introduction}

Accurate measurements of temperature of glasses close to the softening point
($T\sim 1000-2000$ K) require the knowledge of emissivity at the wavelengths
used by the thermal camera, which are typically in the 4-8 $\mu$m (mid-wavelength
infrared, MWIR) and 8-14 $\mu$m (long-wavelength infrared, LWIR) ranges. Because in
the laser processing of glasses, the laser-heated region can be arbitrarily
thin, the emissivity, especially in the MWIR range, can be diminished and
calibration procedures performed on the bulk glass can be invalidated.

Density functional theory (DFT) simulations can provide accurate information
about the infrared response, absorption, and emission of the material. To
account for anharmonism and finite temperatures, the molecular dynamics (MD)
Green-Kubo approach was developed\cite{bader_berne, iftimie, thomas2013}. One
difficulty with this method is that the dipole moment is ill-defined in the
periodic boundary conditions, which are most suited for simulations of glass.
This difficulty was overcome through the development of the modern theory of
polarization\cite{resta1992, king1993theory}. With its use, the infrared
response was previously computed with the Green-Kubo approach in Refs.
\cite{iftimie, thomas2013} via calculation of maximally-localized Wannier
orbitals. Below, we demonstrate another approach which, within real-time
time-dependent DFT (rt-TDDFT), directly computes the change of the macroscopic
polarization as the total current through the simulation cell. We also test
simplified approaches based on charge partitioning schemes.

\section{Materials and Methods}

In this work, our objective is to calculate the emission and absorption of IR
radiation by glass at high temperature close to the softening point.
Additionally, the OH (hydroxyl) emission and absorption band is of particular
interest.

In order to make a comparison with experimental data it useful to calculate the spectral radiance  $L$:
\begin{equation}
L(\omega, T) = \frac{\hbar \omega^3}{4 \pi^3 c^2}\frac{1 - \ee^{-\alpha(\omega) d}}{\ee^{\frac{\hbar \omega}{k_\text{B} T}} - 1},
\end{equation}
where $\alpha$ is the Napierian absorption coefficient and $d$ is the thickness
of the emitting layer (see Appendix \ref{apxa} for relations between different
absorption coefficients). In this work, a hot layer of uniform temperature is
considered for simplicity.

The IR light is absorbed in the material through vibrational excitation of atomic dipoles.
The quantum dipole-dipole autocorrelation function and its Fourier transform
can be approximately computed via the classical autocorrelation
function\cite{bader_berne}:
\begin{eqnarray}
\int_{-\infty}^{\infty} \dd t \eexp^{-i \omega t} \langle \frac{1}{2}\left[\hat{\vmu}(0),  \hat{\vmu}(t) \right]_+ \rangle \nonumber \\
= \frac{\beta \hbar \omega}{2} \coth(\frac{\beta \hbar \omega}{2}) \int_{-\infty}^{\infty} \dd t\ \eexp^{-i \omega t} \langle \vmu(0) \cdot  \vmu(t) \rangle_{\text{cl}},
\label{eq:cl}
\end{eqnarray}
where $[]_+$ denotes the anticommutator, $\vmu$  is the total dipole moment of the system, $\hat{\vmu}$ is the dipole moment operator and $\beta=1/k_\text{B} T$.

The imaginary part $\varepsilon_2(\omega)$ of the dielectric function can be
found via Green-Kubo relation for the dipole or dipole time-derivative autocorrelation function\cite{iftimie}:
\begin{multline}
\varepsilon_2(\omega) =  \frac{\beta}{6 V \omega \varepsilon_0} \int_{-\infty}^{\infty} \dd t\ \eexp^{-i \omega t} \langle \dot{\vmu}(0) \cdot  \dot{\vmu}(t) \rangle_{\text{cl}},
\label{eq:imeps}
\end{multline}
where $V$ is the system's volume and the dot indicates the time-derivative. The time-derivative form of the relation is chosen for numerical efficiency.

The real part of the dielectric function $\varepsilon_1(\omega)$ is
obtained from the Kramers-Kr\"{o}nig relation:
\begin{equation}
\varepsilon_1(\omega) - 1 = \frac{2}{\pi} \Pcal \int_0^\infty \frac{\omega' \varepsilon_2(\omega')}{\omega'^2 - \omega^2} \dd \omega',
\label{eq:KK}
\end{equation}
with $\Pcal$ indicating Cauchy principal value. Eq. (\ref{eq:KK}) can
either be evaluated via direct numerical integration, or by taking two successive Fourier transforms\cite{frohlich,iftimie}:
\begin{equation}
\varepsilon_1(\omega) - 1 = \frac{2}{\pi} \int_0^\infty \int_0^\infty \varepsilon_2(\omega') \sin(\omega' \tau) \cos(\omega \tau) \dd \omega' \dd\tau
\end{equation}

In the periodic boundary conditions used in the calculations below the dipole moment is
not defined.  One way to deal with this problem is to use localized Wannier functions
\cite{resta_review, iftimie, thomas2013}. However, this requires finding localized Wannier functions at
each MD time step. Another approach, which we propose here, is to directly calculate
the autocorrelation function for the time derivative of the macroscopic polarization $\mathbf{P}(t)$,
which is the total current through the simulation cell\cite{resta1992, king1993theory, resta_review}:
\begin{equation}
\mathbf{P}(t) = \frac{1}{V_{\text{cell}}} \int_0^t \mathbf{j}_{\text{cell}}(t') dt'.
\end{equation}
$\mathbf{j}_{\text{cell}}$ can be calculated either directly in real-time time-dependent
DFT (rt-TDDFT, Method 1) or by calculation of the partial charges on atoms and multiplying them
by the atomic velocity vectors (Method 2).

\paragraph*{Numerical simulations.}
We used the SIESTA\cite{siesta02} DFT code with certain modifications
implemented by us, including rt-TDDFT\cite{kolesov2015real}.  We used the PBE
exchange-correlation  functional\cite{pbe} and the double-$\zeta$+polarization
(DZP) basis set. We found that this higher-accuracy basis is  necessary to
correctly reproduce the frequency of the OH stretch vibration.

A fused-silica simulation cell was prepared by simulated annealing at
constant volume for 20 ps. The resulting simulation cell contained 95 atoms
including two hydroxyl groups. We used a time step of 0.3 fs in all our
standard (adiabatic) molecular dynamics simulations.  Such a reduced time step is necessary
to capture motion of the hydrogen atom. In the rt-TDDFT calculations we used a time step
of 1 a.u. ($\approx 24$ as). In order to obtain the realistic concentration of
OH we  also calculated the absorption spectrum for the fused-silica cell with
no OHs  and then calculated the weighted average of the two spectra with weights
made to match the experimental OH concentration.

We used the Hirshfeld and Bader charge partitioning methods for the partial
charge-based method (Method 2).  We also tested the Voronoi deformation density
charge partitioning scheme\cite{voronoi}, however the results of this scheme were very
similar to those obtained with the Hirshfeld method, so only the latter is
reported here. Bader charges were calculated with \emph{Bader}
code\cite{badercode}.


\section{Results and discussion}

\subsection{Fused silica.}

\begin{figure*}[ht]
\includegraphics[width=0.8\textwidth]{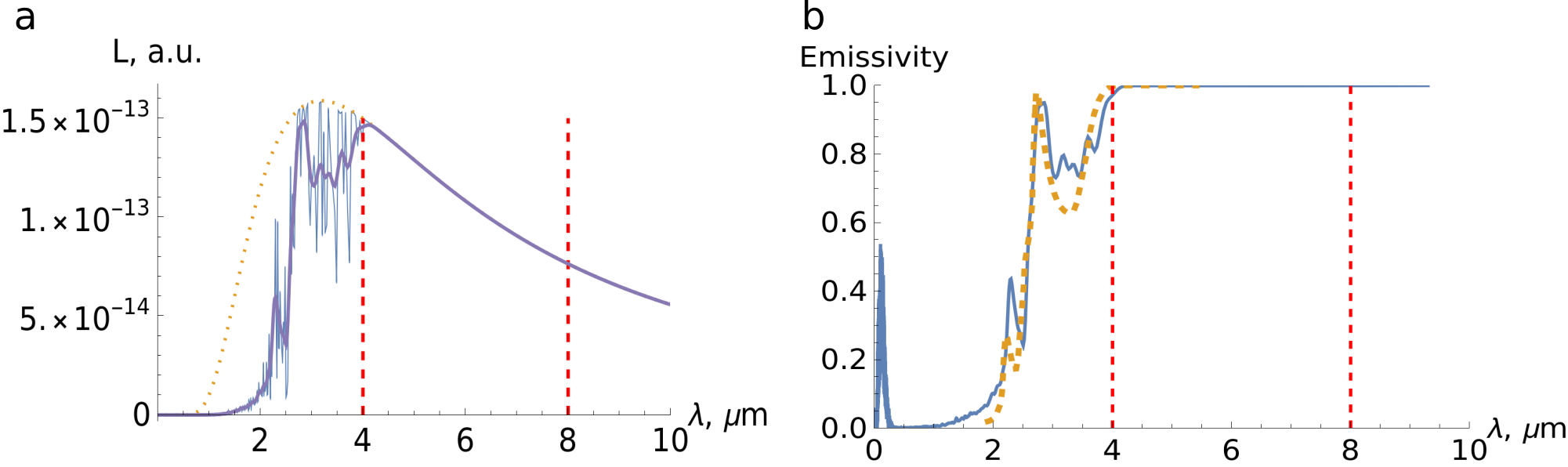}
\caption{ Method 1: The calculated emission spectra (a) and the spectral emissivity (b), for a 6mm-thick,  0.15 weight-\% OH fused silica sample at $T=1600$ K. In (a) the orange dotted line shows the black-body emission spectrum, the thin blue line shows the calculated emission spectrum while the purple line shows the smoothed version of the calculated spectrum. The latter was used to calculate (b). In (b) the solid blue line shows the calculated spectral emissivity and the dashed orange line shows the experimental curve from Ref.\cite{dvurechensky}. In both plots the red dashed lines mark the wavelengths  range between 4 $\mu$m and 8 $\mu$m often used by thermal cameras.}
\label{fig:emit1}
\end{figure*}

\begin{figure*}[ht]
\includegraphics[width=0.8\textwidth]{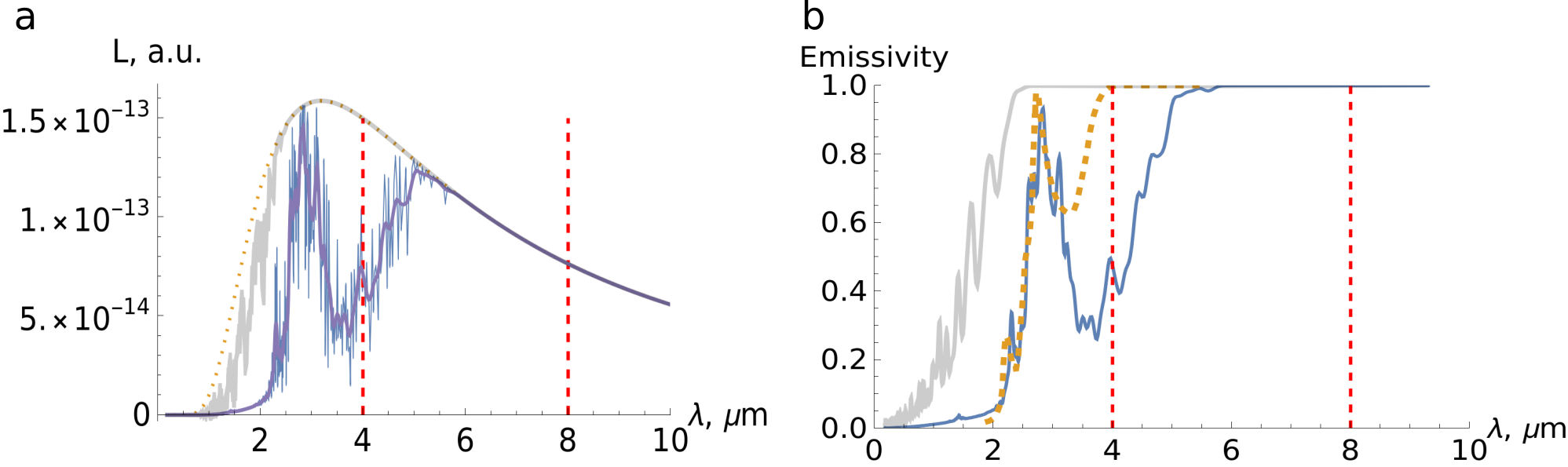}
\caption{ Method 2: The calculated emission spectra (a) and the spectral emissivity (b), for a 6mm-thick,  0.15 weight-\% OH fused silica sample at $T=1600$ K. In (a) the orange dotted line shows the black-body emission spectrum, the thin blue line shows the calculated emission spectrum (Hirshfeld charge method) while the purple line shows the smoothed version of the calculated spectrum. The thick gray line shows the spectrum calculated with Bader charges.  In (b) the solid blue line (Hirshfeld charge method, smoothed) and thick gray line (Bader charge method) show the calculated spectral emissivity. The dashed orange line shows the experimental curve from Ref.\cite{dvurechensky}.}
\label{fig:emit2}
\end{figure*}


\begin{table*}[ht]
\begin{tabular}{l|c|c|c|c}
 & ~~Ref\cite{williams}~~ & ~~Ref\cite{dvurechensky}~~ & rt-TDDFT (this work, Method 1) & DFT (this work, Method 2) \tabularnewline
\hline
$\alpha$, cm$^{-1}$ ($10^3$ weight--ppm OH)& 12.83                    & 9.87            & 14.45              & 13.66                     \tabularnewline
$\epsilon_{10}$, L cm$^{-1}$/mol           & 43                       & 33.1            & 48.44              & 45.78                     \tabularnewline
$\sigma$,  $10^{-19}$ cm$^2$            & 1.64                        & 1.27            & 1.85               & 1.75                      \tabularnewline
Peak wavelength, $\mu$m                    & 2.73                     & 2.72            & 2.79               & 2.85                      \tabularnewline
\end{tabular}
\caption{ Experimental and theoretical OH-band peak absorption coefficients for fused silica. }
\label{tababs}
\end{table*}

Table \ref{tababs} presents the absorption coefficients for the peak OH band
absorption. The Hirshfeld charge-based method (Method 2) and the rt-TDDFT method
(Method 1) both are in good agreement with the experimental measurements.
Both methods overestimate the absorption somewhat, but in both cases it is quite close to the
measurement by Williams \emph{et al.}\cite{williams}. Note that Method 1 derives the
absorption spectrum from the calculation, which is accurate within the
simulation, while Method 2 is an approximate way of extracting absorption from
the simulation. However it utilizes more trajectory data points due to the low cost
of each MD time step in adiabatic DFT simulations.

Note that the effect of peak broadening due to configurational disorder is not
explored in this work; we calculate the absorption exactly at the peak. Thus,
there is some uncertainty associated with the calculated values presented in
Table \ref{tababs}. Ideally, we would conduct many simulations on different
SiO$_2$ + OH configurations and then average the resulting spectra.

Figs. \ref{fig:emit1} (Method 1) and  \ref{fig:emit2} (Method 2) present a
more integral way of comparing the results to the experiment.  Fig. \ref{fig:emit1}a shows
the spectral radiance of the 6mm 0.15 weight-\% OH fused silica sample at
$T=1600$ K. In order to average out the discrete peaks resulting from the
finite-size simulation cell (the thin blue line on Fig. \ref{fig:emit1}a)  we applied
low-pass filter (the thick purple line). This smoothed-out version of the spectral
radiance was then used to calculate the emissivity (Fig. \ref{fig:emit1}b). 
The experimental emissivity curve ($T=1675$ K) measured by Dvurenchesky \emph{et al.}
\cite{dvurechensky} is shown as the dashed orange line in Fig. \ref{fig:emit1}b.
The agreement between the simulation and the experiment is good.  The OH-peak
is slightly red-shifted, but otherwise it is where it should be and has a correct
width. The emissivity is dropping off from the black-body value at $\sim 4\
\mu$m both in the experiment and the simulation.  The overlap between OH-peak
and the tail of the black-body emission is also in agreement.  The satellite
(combination) peak at$~\sim2.3\ \mu$m is reproduced well. It is also
slightly red-shifted, but its intensity and width agree with the experiment
within the uncertainty of the simulation (the thin blue line in Fig.
\ref{fig:emit1}a).

The simulation results show a peak in the emissivity at low wavelengths.
It corresponds to the electronic excitations, which are also captured by the
rt-TDDFT method. It is indistinguishable from zero on the absolute scale (Fig.
\ref{fig:emit1}a) and is not seen in the experiment, as it is out
of the spectrometer range.

The results of the calculation with Method 2 and Hirshfeld charges for a
6mm-thick sample are shown in Fig.  \ref{fig:emit2}. The OH-peak and the
combination mode peak at 2.3 $\mu$m are captured well with this method. The
OH-overtone mode at $\sim 1.4$ $\mu$m is clearly seen in Fig. \ref{fig:emit2}b.
However the drop-off from the black-body behavior starts at too large a
wavelength $\sim 5.5\ \mu$m, resulting in the local minimum  of the emissivity
(0.7) at 4 $\mu$m.  Evidently, bulk silica dipoles are not represented well
with this method, resulting in the overall poor emission and absorption
prediction. This is possibly due to the underestimation of partial charges on
ions in more covalent Si-O bonds, but correct charge assignments in polar O-H
bond with the Hirshfeld charge partitioning scheme.

Fig. \ref{fig:emit2} also shows the spectrum and spectral emissivity calculated
with Bader charges. Typically, Bader charge partitioning results in
near-integer charges consistent with the oxidation state of an ion. In the
simulations presented here, Bader charges significantly overestimate the actual
effective charges, which results in the incorrect, nearly black-body emission
spectrum.

This assessment of Hirshfeld, Voronoi and Bader charges broadly agrees
with the conclusions of Ref. \cite{voronoi}. 

\begin{figure*}[ht]
\includegraphics[width=0.8\textwidth]{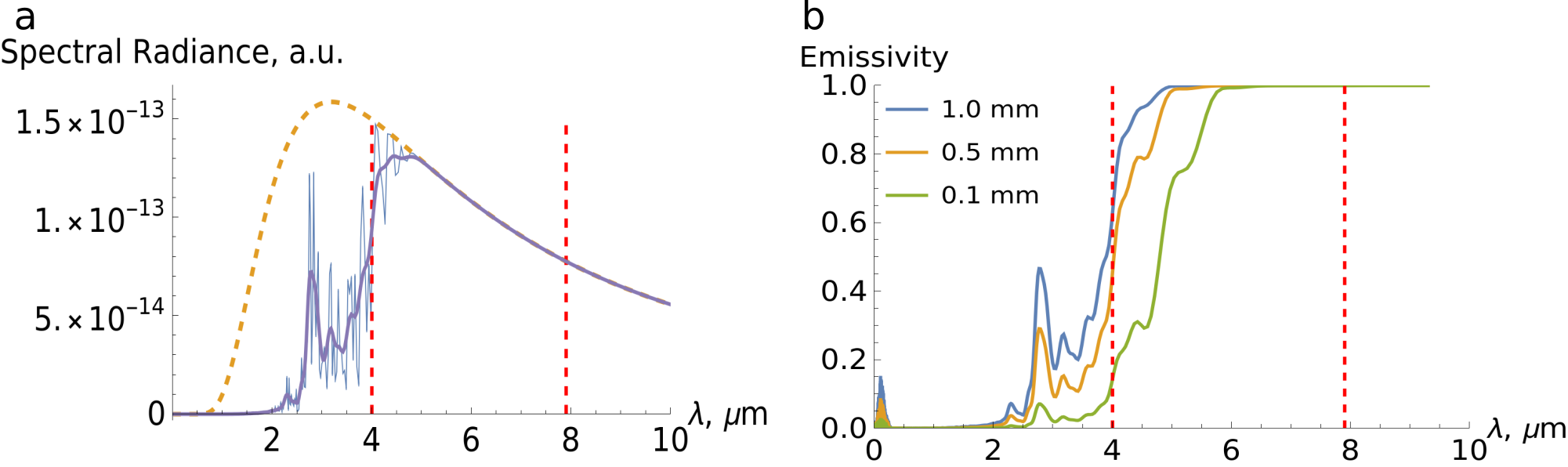}
\caption{ The effect of the heated region thickness on the emissivity for the 1000-ppm OH-content HPFS glass. The calculated emission spectra for  the 1 mm-thick  heated region (a) and the spectral emissivity (b) for the 1 mm (blue curve), 0.5 mm (orange) and 0.1 mm (green)-thick heated regions. In (a) the orange dotted line shows the black-body emission spectrum, the thin blue line shows the calculated emission spectrum, and the purple line shows the smoothed version of the calculated spectrum. In both plots the red dashed lines mark the wavelengths  range between 4 $\mu$m and 8 $\mu$m often used by thermal cameras.
}
\label{fig:thick}
\end{figure*}

One of the goals of this calculation is to understand thermal-camera measurements in
laser processing of glass. In the laser processing the heated region does
not necessarily extend through the whole sample and can be arbitrarily thin.
The effect of the heated-region thickness is explored in Fig.
\ref{fig:thick} for 1000-ppm OH fused silica (similar in OH-content to the
Corning-made HPFS).

The temperature measurement at 4 $\mu$m would be unreliable even for the 1 mm
sample, where the 4 $\mu$m line falls right onto the step-like  transition  between
the black-body and the structured radiation. It should be noted that the calculation
uncertainty is very high at 4 $\mu$m (as seen from the non-smoothed calculation curve
depicted by the thin blue line in Fig. \ref{fig:thick}a). The emissivity at
4 $\mu$m is significantly lower than 1 for the 0.5 mm and 0.1 mm thicknesses,
approximately 0.45 and 0.15, respectively. The temperature measurement at 8
$\mu$m would be reliable in all cases.

\subsubsection*{OH-stretch overtone}

It is interesting to see that in the emission spectra calculated with Hirshfeld
charges, the OH-vibration overtone peak at about 1.4 $\mu$m is easily seen. 
Since this peak is of particular importance in optical fibers, it was investigated
further. 

\begin{figure*}[ht]
\includegraphics[width=0.8\textwidth]{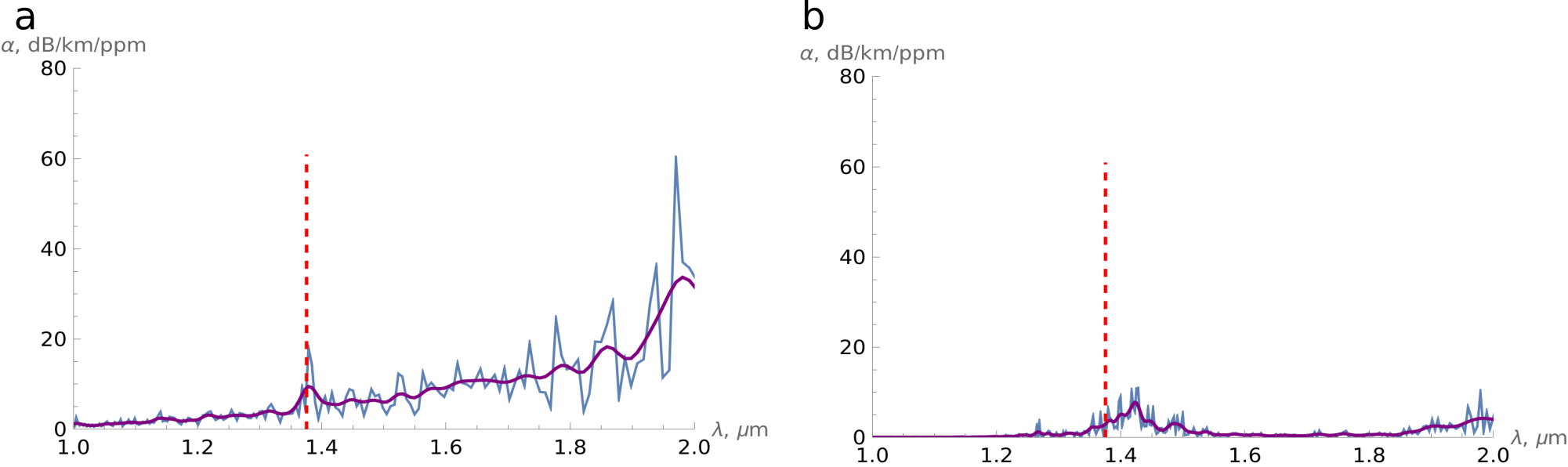}
\caption{ The absorption coefficient in the conventional dB/km/ppm units around the 1.4 $\mu$m wavelength ($T=1600$ K). (a) Calculated with Method 1 and (b) with Method 2. In both plots blue curve is the calculated absorption coefficient, thick purple line is the smoothed version and red dashed line indicates OH-overtone peak position and height at room temperature\cite{humbach1996}.    }
\label{fig:ov}
\end{figure*}

Fig. \ref{fig:ov} presents the absorption coefficient at $T=1600$ K in the
customary units of dB/km/ppm calculated with both methods. The overtone peak is
seen with both methods, although with Method 1 it is barely distinguishable from
the background absorption, possibly resulting from the thermal broadening of the
2.3 $\mu$m combination peak. With both methods the height of the peak is
significantly below the room-temperature value\cite{humbach1996}. With Method 1,
however the peak height could be consistent with the temperature-dependence trend
observed in the Ref. \cite{yu2016}, where the peak intensity at $T=500$ K was
measured to be significantly below the room-temperature value. With Method 2, the
peak is more distinct, but it is significantly red-shifted and lower in
intensity than the room-temperature value.

To clarify this further, the absorption spectrum at $T=300$ K  was calculated with
Method 1. The overtone peak was present, but it was two orders of magnitude below the
experimental value. Clearly for such a high frequency ($\lambda=1.4\ \mu$m,
$\omega=0.9\text{~eV}=10277\text{ K}$), the assumptions behind the
quantum-classical equivalence Eq. (\ref{eq:cl}) are not satisfied. In
this classical-ion simulation the temperature on energy scale was much lower than the frequency of
the OH-stretch mode, resulting in only a slight excitation of the mode and
limited exploration of the anharmonic part of the potential. This is in
contrast to a quantum oscillator, which is half-excited in the zeroth state.

Therefore, the Green-Kubo-based method is unreliable for high-frequency vibrational
overtones due to the failure of quantum-classical equivalence for high-frequency
vibrations, which are in the quantum zero-point regime.

\subsection{Borofloat}

\begin{figure*}[ht]
\includegraphics[width=0.8\textwidth]{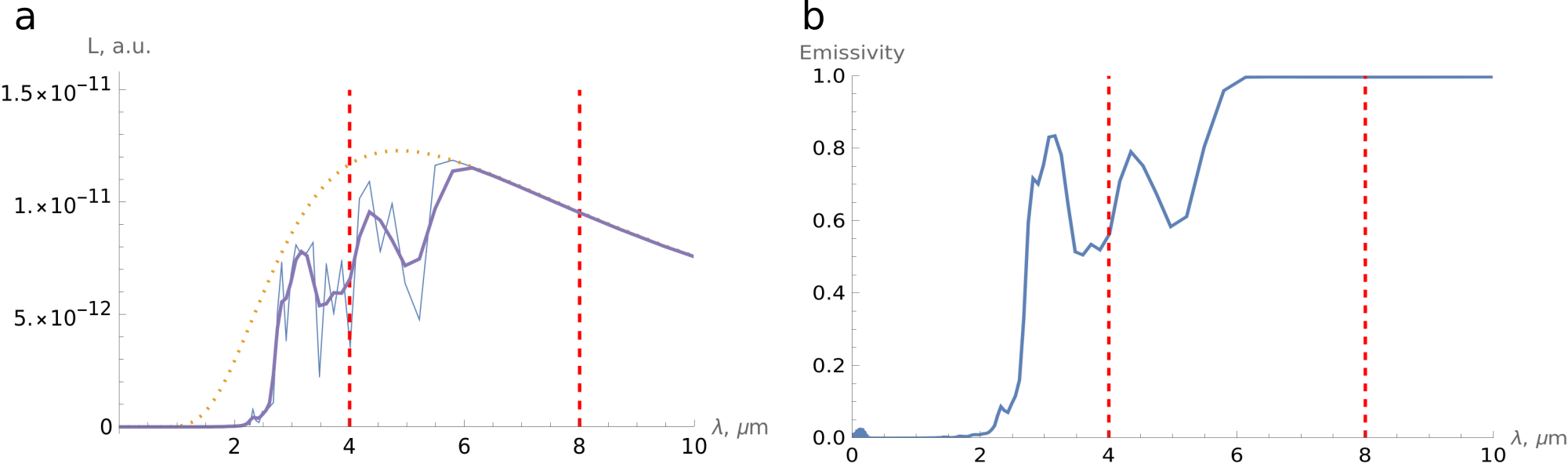}
\caption{ The calculated emission spectra (a) and the spectral emissivity (b), for a 100 $\mu$m-thick,  1 weight-\% OH Borofloat sample at $T=1050$ K. In (a) the  orange dotted line shows the black-body emission spectrum, the thin blue line shows the calculated emission spectrum and the purple line shows the smoothed version of the calculated spectrum. The latter was used to calculate (b). In (b) the solid blue line shows the calculated spectral emissivity.}
\label{fig:boro1}
\end{figure*}

\begin{figure}
\includegraphics[width=0.4\textwidth]{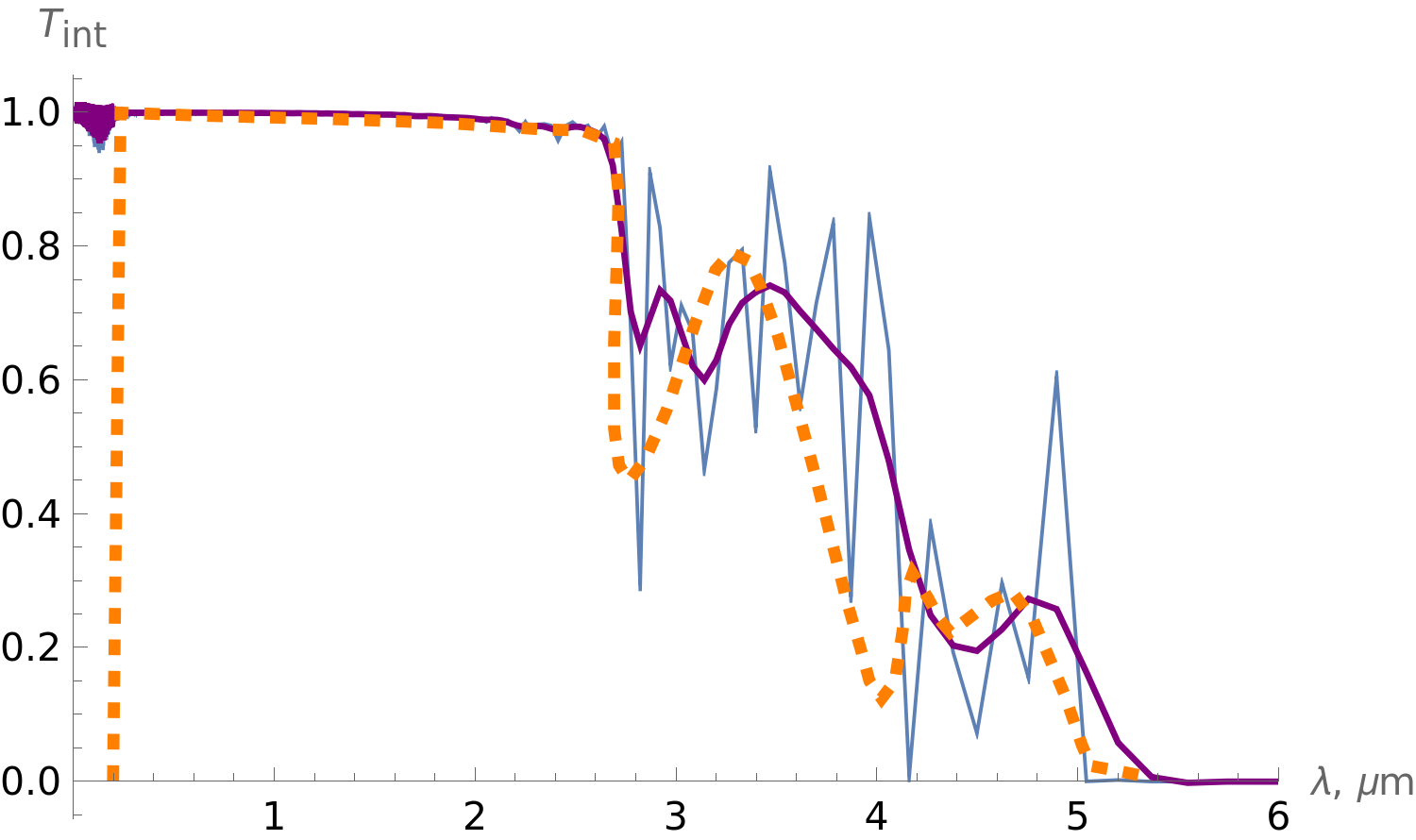}
\caption{ The transmission spectrum of 0.2 mm-thick  BOROFLOAT\R\ 33 at room temperature. The orange dashed curve shows the experimental transmission spectrum from the BOROFLOAT\R\ 33 specification webpage (digitized and normalized to be 1 at the maximum of transmission). Purple thick line shows the smoothed version of the calculated transmission spectrum at $T=300$ K and 2000 ppm OH-content, while the original non-smoothed spectrum is shown with the thin blue line. }
\label{fig:boro2}
\end{figure}

To explore the technique developed in this work on more complex glass we
created the simulation cell for the Borofloat glass. The composition of Borofloat was taken
from the BOROFLOAT\R\ 33 specification web page and is, in mol-\%, 81\% SiO$_2$, 13\%
 B$_2$O$_3$, 4\% Na$_2$O/K$_2$O, 3\% Al$_2$O$_3$. We added 1\% of OH
molecules into the composition and adjusted the content of the other oxides
accordingly.  The resulting simulation cell had about 200 atoms. We
applied rt-TDDFT at $T=1050$ K to calculate the IR absorption and emission.  The
results for the 0.1 mm-thick heated region are presented in Fig.
\ref{fig:boro1}. The 0.1 mm thickness was chosen so that the spectrum structure
around 4 $\mu$m would be visible.

The main OH-peak and the 2.3 $\mu$m combination peak are again clearly
distinguishable.  At room temperature the transmission spectrum for 2000 ppm
OH-content matches reasonably well given the uncertainty in the composition and
finite size effects to the transmission spectrum measured by Schott (Fig.
\ref{fig:boro2}).

\section{Conclusions}

The Green-Kubo approach, in combination with rt-TDDFT, was used to calculate the
absorption and emission of glasses at high-temperature, close to the softening
point. The autocorrelation functions of the total current were transformed to
obtain the IR response at a given temperature.  Good agreement with the
experimental measurements was observed for fused silica. The OH-band,  with its
satellite peaks, which is of particular importance for this study, was well
reproduced.

A multi-component glass, BOROFLOAT\R\ 33, was also considered in this work. The
simulation reproduced the transmission spectrum provided in the BOROFLOAT\R\ 33
specification well, given the uncertainty in the exact composition and
finite size effects.

All simulations demonstrated the risk associated with temperature measurements
of the laser-heated spots using the IR camera that operates below 6 $\mu$m
wavelengths. For these wavelengths, measurements calibrated on thick glass
samples may not accurately reflect the temperature of the thin laser-heated
region due to its lower emissivity.

For very high frequency vibrational bands, in particular OH overtone band, the
method was not able to accurately reproduce the peak intensity, although the
peak position was reasonably accurate. This is because for high frequency
vibrations, even at temperatures around the softening point,  the vibrational
state is in the quantum zero-point regime which is not reproduced well in the
quantum-to-classical mapping used in the classical MD-based Green-Kubo approach
used in this work.

When we attempted to simplify the calculation of the total current by using
Hirshfeld or Voronoi partial charges, the method was not able to accurately
reproduce the experimental IR emission. Interestingly, while the OH-band was
relatively well reproduced, the emission related to the top of the SiO$_2$ band
was poorly reproduced. We deduce that Hirshfeld and Voronoi partial charge
methods work well for highly polar bonds, such as OH-bonds, but fail for less
polar bonds, such as Si-O. Bader charge partitioning scheme resulted in 
overestimation of the emission intensities and nearly black-body emission that
drowned all of the structured signal. Thus, IR emission and absorption spectra serve as
useful benchmark to assess the applicability of the charge partitioning schemes.

\section*{Acknowledgements}

The author is grateful to Dr. Alex Streltsov for introduction to the problem and
many fruitful discussions, and to Dr. Douglas Allan for thorough reading and providing feedback on the
manuscript.

\appendix\section{Relations between different optical constants\label{apxa}}

The average rate of absorption of energy $E$ per light absorber  is given by
\begin{equation}
    \frac{\partial{E}}{\partial t} = \sigma I,
\end{equation}
where $I$ is the light intensity (power per unit area) and 
$\sigma$ is the absorption cross-section. $\sigma$ is related to the molar absorption coefficient $\epsilon$ through
\begin{equation}
    \sigma = \epsilon / N_{\text{A}}, 
\end{equation}
where $N_{\text{A}}$ is the Avogadro constant. 

The
absorption coefficient $\alpha$ and the molar absorption coefficient $\epsilon$ are related through
the molar concentration  $q$ of the absorbing species:
\begin{equation}
    \alpha = q\cdot \epsilon 
\end{equation}

The Napierian  $\alpha,~\epsilon$ (base--e) and the
decadic $\alpha_{10},~\epsilon_{10}$ (base--10) absorption coefficients
are related by the $\log$-factor:
\begin{eqnarray}
    \alpha_{10} &=& \alpha   / \log{10}  \nonumber \\
    \epsilon_{10}&=& \epsilon / \log{10},
\end{eqnarray}
where $\log$ indicates the Napierian logarithm.

The Napierian $\alpha$, measured in cm$^{-1}$, is converted to dB/km/ppm units with
\begin{equation}
	\alpha_{\text{dB}} = \alpha 10^6 \log_{10}{\ee} / q_{\text{ppm}},
\end{equation}
where $q_{\text{ppm}}$ is the concentration  of the absorbing species in ppm.

Frequency-dependent absorption coefficient $\alpha(\omega)$ is related to the
complex refractive index  $\hat{n}(\omega)$ and complex dielectric function
$\varepsilon(\omega)$:
\begin{eqnarray}
\varepsilon(\omega) &=& \varepsilon_1(\omega) + i \varepsilon_2(\omega) \nonumber\\
\hat{n}(\omega)     &=& n(\omega) + i k(\omega) \nonumber\\
\varepsilon(\omega) &=& \hat{n}^2 \nonumber\\ 
\alpha(\omega)      &=& \frac{2 \omega}{c} k(\omega) \nonumber\\
\varepsilon_2(\omega) &=& \frac{c}{\omega} \alpha(\omega) n(\omega),
\label{enka}
\end{eqnarray}
where $\varepsilon_1$ and $\varepsilon_2$ are real and imaginary parts of the
complex dielectric function, and $n(\omega)$ and $k(\omega)$ are real and
imaginary part of the complex refractive index. (Magnetic permeability $\mu$ is
assumed to be unity.)

\clearpage

\section*{References}
\bibliographystyle{naturemag}
\bibliography{refs_nourl}

\end{document}